\documentclass[11pt,twoside]{article}

\usepackage{asp2014}

\aspSuppressVolSlug
\resetcounters

\begin{document}

\title{The New Architecture of the Online Observation Quality System for the ASTRI Mini-Array Project}

\author{N. Parmiggiani,$^1$ A. Bulgarelli,$^1$ L. Castaldini,$^1$ V. Fioretti,$^1$ I. Abu,$^1$ M. Capalbi,$^2$ O. Catalano,$^2$ V. Conforti,$^1$ M. Corpora,$^2$ A. Di Piano,$^{1,3}$ R. Falco,$^1$ M. Fiori,$^{4}$ F. Gianotti,$^1$ S. Iovenitti,$^5$, F. Lucarelli$^{6,7}$ M. C. Maccarone,$^2$ T. Mineo,$^2$ D. Mollica,$^2$ S. Lombardi,$^{6,7}$ G. Panebianco,$^1$ V. Pastore,$^1$, A. Rizzo$^8$, F. Russo$^1$, P. Sangiorgi,$^2$ S. Scuderi,$^{9}$ G. Tosti,$^{10}$ L. Zampieri,$^4$ and for the ASTRI Project$^{11}$}
\affil{$^1$INAF/OAS Bologna, Via P. Gobetti 93/3, 40129 Bologna, Italy; \email{nicolo.parmiggiani@gmail.it}}
\affil{$^2$INAF/IASF Palermo, Via Ugo La Malfa 153, I-90146 Palermo, Italy}
\affil{$^3$Universit\`{a} degli Studi di Modena e Reggio Emilia, DIEF, Via Pietro Vivarelli 10, 41125 Modena, Italy.}
\affil{$^4$INAF/OA Padova, Vicolo Osservatorio 5, I-35122 Padova, Italy}
\affil{$^5$INAF/OA Brera, via Brera 28, I-20121 Milano, Italy}
\affil{$^6$INAF/OAR Roma, Via di Frascati 33, I-00078, Monte Porzio Catone, Roma, Italy}
\affil{$^7$ASI/SSDC Roma, Via del Politecnico snc, I-00133 Roma, Italy}
\affil{$^8$INAF/OA Catania, Via Santa Sofia 78, 95123 Catania, Italy}
\affil{$^{9}$INAF/IASF Milano, Via Alfonso Corti 12, I-20133 Milano, Italy}
\affil{$^{10}$Universit\`{a} degli Studi di Perugia, Dip.to di Fisica e Geologia, Via A. Pascoli, I-06123 Perugia, Italy}
\affil{$^{11}$\protect\url{http://www.astri.inaf.it/en/library/}}

\paperauthor{Nicol\`{o}~Parmiggiani}{nicolo.parmiggiani@inaf.it}{0000-0002-4535-5329}{INAF}{OAS}{Bologna}{BO}{40129}{Italy}
\paperauthor{Andrea~Bulgarelli}{andrea.bulgarelli@inaf.it}{0000-0001-6347-0649}{INAF}{OAS}{Bologna}{BO}{40129}{Italy}
\paperauthor{Valentina~Fioretti}{valentina.fioretti@inaf.it}{0000-0002-6082-5384}{INAF}{OAS}{Bologna}{BO}{40129}{Italy}
\paperauthor{Ismam~Abu}{ismam.abu@inaf.it}{0009-0001-7973-8192}{INAF}{OAS}{Bologna}{BO}{40129}{Italy}
\paperauthor{Milvia~Capalbi}{milvia.capalbi@inaf.it}{0000-0002-9558-2394}{INAF}{IASF}{Palermo}{PA}{I-90146}{Italy}
\paperauthor{Osvaldo~Catalano}{osvaldo.catalano@inaf.it}{0000-0002-9554-4128}{INAF}{IASF}{Palermo}{PA}{I-90146}{Italy}
\paperauthor{Vito~Conforti}{vito.conforti@inaf.it}{0000-0002-0007-3520}{INAF}{OAS}{Bologna}{BO}{40129}{Italy}
\paperauthor{Mattia~Corpora}{mattia.corpora@inaf.it}{0000-0002-5374-157X}{INAF}{IASF}{Palermo}{PA}{I-90146}{Italy}
\paperauthor{Ambra~Di~Piano}{ambra.dipiano@inaf.it}{0000-0002-9894-7491}{INAF}{OAS}{Bologna}{BO}{40129}{Italy}
\paperauthor{Riccardo~Falco}{riccardo.falco@inaf.it}{0009-0004-1676-7596}{INAF}{OAS}{Bologna}{BO}{40129}{Italy}
\paperauthor{Michele~Fiori}{michele.fiori@inaf.it}{0000-0002-7352-6818}{INAF}{OAS}{Padova}{PD}{I-34122}{Italy}
\paperauthor{Fulvio~Gianotti}{fulvio.gianotti@inaf.it}{0000-0003-4666-119X}{INAF}{OAS}{Bologna}{BO}{40129}{Italia}
\paperauthor{Simone~Iovenitti}{simone.iovenitti@inaf.it}{0000-0002-2581-9528}{INAF}{OA}{Brera}{MI}{I-20121}{Italy}
\paperauthor{Fabrizio~Lucarelli}{fabrizio.lucarelli@inaf.it}{0000-0002-6311-764X}{INAF}{OAR}{Roma}{RO}{I-00078}{Italy}
\paperauthor{Maria~Concetta~Maccarone}{cettina.maccarone@inaf.it}{0000-0001-8722-0361}{INAF}{IASF}{Palermo}{PA}{I-90146}{Italy}
\paperauthor{Teresa~Mineo}{teresa.mineo@inaf.it}{0000-0002-4931-8445}{INAF}{IASF}{Palermo}{PA}{I-90146}{Italy}
\paperauthor{Davide~Mollica}{davide.mollica@inaf.it}{0000-0003-4862-9927}{INAF}{IASF}{Palermo}{PA}{I-90146}{Italy}
\paperauthor{Saverio~Lombardi}{saverio.lombardi@inaf.it}{0000-0002-6336-865X}{INAF}{OAR}{Roma}{RO}{I-00078}{Italy}
\paperauthor{Gabriele~Panebianco}{gabriele.panebianco@inaf.it}{0000-0002-3410-8613}{INAF}{OAS}{Bologna}{BO}{40129}{Italy}
\paperauthor{Valerio~Pastore}{valerio.pastore@inaf.it}{0000-0002-4776-5890}{INAF}{OAS}{Bologna}{BO}{40129}{Italy}
\paperauthor{Alessandro~Rizzo}{alessandro.rizzo@inaf.it}{0009-0003-4341-2988}{INAF}{OA}{Catania}{CT}{95123}{Italy}
\paperauthor{Federico~Russo}{federico.russo@inaf.it}{0000-0002-3476-0839}{INAF}{OAS}{Bologna}{BO}{40129}{Italy}
\paperauthor{Pierluca~Sangiorgi}{pierluca.sangiorgi@inaf.it}{0000-0001-8138-9289}{INAF}{IASF}{Palermo}{PA}{I-90146}{Italy}
\paperauthor{Salvatore~Scuderi}{pierluca.sangiorgi@inaf.it}{0000-0002-8637-2109}{INAF}{IASF}{Milano}{MI}{I-20133}{Italy}
\paperauthor{Gino~Tosti}{gino.tosti@inaf.it}{0000-0002-0839-4126}{Universit\`{a} degli studi di Perugia}{Dipartimento di Fisica e Geologia}{Perugia}{Itali}{I-06123}{Italy}
\paperauthor{Luca~Zampieri}{luca.zampieri@inaf.it}{0000-0002-6516-1329}{INAF}{OAS}{Padova}{PD}{I-34122}{Italy}



\begin{abstract}
The ASTRI Mini-Array is an international collaboration led by the Italian National Institute for Astrophysics. The project aims to construct and operate an array of nine Imaging Atmospheric Cherenkov Telescopes to study gamma-ray sources at very high energy (TeV) and perform stellar intensity interferometry observations. We describe the updated Online Observation Quality System (OOQS) software architecture. The OOQS is one of the subsystems of the Supervisory Control and Data Acquisition (SCADA) system. It aims to execute real-time data quality checks on the data acquired by the Cherenkov cameras and intensity interferometry instruments and provide feedback to both SCADA and the Operator about abnormal conditions detected.  The data quality results are stored in the Quality Archive for further investigation and sent to the Operator Human Machine Interface (HMI) through Kafka.
\end{abstract}



\section{Introduction}

The ASTRI Mini-Array \citep{SCUDERI202252} is a project led by the Italian National Institute for Astrophysics (INAF) that aims to construct and operate nine Cherenkov telescopes to study gamma-ray sources in the TeV energy range and perform intensity interferometry observations. The telescopes are equipped with a Stellar Intensity Interferometry Instrument (SI3) \citep{Zampieri2022} that can be deployed during the SI3 observations. The ASTRI Mini-Array is under installation at the Teide Astronomical Observatory, operated by the Instituto de Astrofisica De Canarias (IAC), on Mount Teide (2400 m a.s.l.) in Tenerife (Canary Islands, Spain). INAF has a host agreement with the IAC to operate the array. 

The ASTRI Mini-Array will be operated remotely, and no human presence is foreseen on-site during observations. A data center was installed on-site to provide the computing power required by the software systems that operate the telescopes and manage the data acquired. The Supervisory Control and Data Acquisition (SCADA,\citet{10.1117/1.JATIS.10.1.017001}) is the system that manages the startup, shutdown, configuration, and control of all site assemblies and sub-systems to collect monitoring points, manage alarms raised by any assembly, check the health status of all systems, and acquire scientific data. All SCADA sub-systems are deployed in the on-site data center to have a reliable and fast connection with the telescopes and other assemblies. The Online Observation Quality System (OOQS,\citet{10.1117/12.2629278},\citet{2024arXiv240402075C}) is part of the SCADA system and is interfaced with other SCADA sub-systems. The OOQS is a software system that aims to execute data quality checks during the observations performed by the ASTRI Mini-Array telescopes. The abnormal conditions detected by the OOQS are used to execute automated corrective actions to optimize the telescopes' duty cycle and prevent data acquisition when not all quality requirements are satisfied. The quality check thresholds are configurable and will be optimized during the commissioning phase.

\section{OOQS New Software Architecture}

\begin{figure*}[!htb]
	\centering
	  \includegraphics[width=1.0\textwidth]{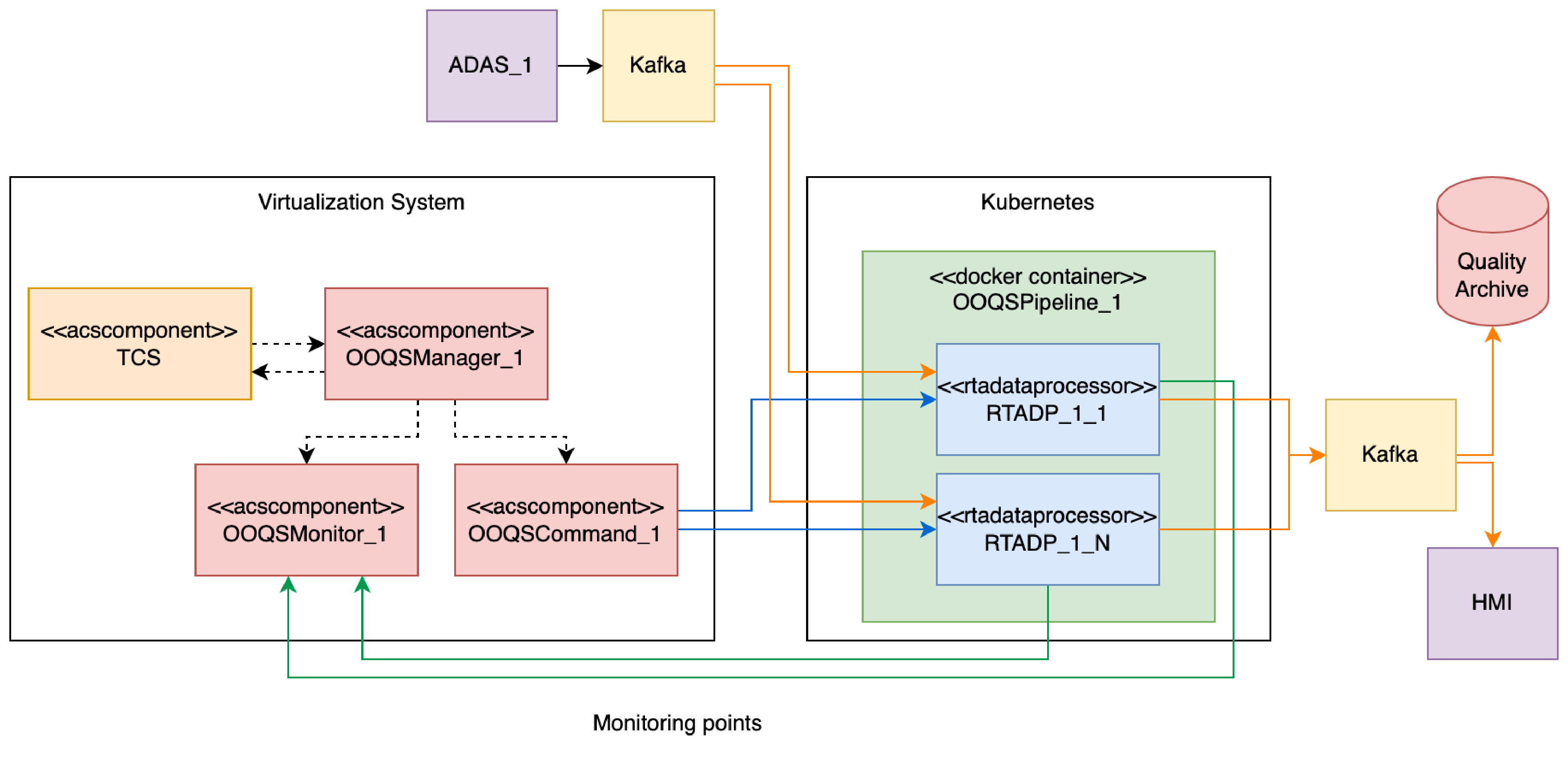}
	\caption{Overview of the OOQS software architecture.}
	\label{fig:architecture}
\end{figure*}

Figure \ref{fig:architecture} shows an overview of the new OOQS software architecture. The OOQS will be deployed in the on-site data center ICT \citep{10.1117/12.3018973} to have a direct connection with the Array Data Acquisition System (ADAS,\citet{10.1117/12.2626600}) from which the OOQS receives the packets acquired by the Cherenkov Cameras through the Kafka event streaming framework\footnote{https://kafka.apache.org/} (orange lines). The OOQS software architecture is composed of two main parts: the management elements, implemented as Alma Common Software\footnote{https://www.eso.org/projects/alma/develop/acs/} (ACS) components, and the analysis pipeline. The former is deployed in the on-site virtualization system while the latter is deployed as Docker\footnote{https://www.docker.com/} container in a Kubernetes\footnote{https://kubernetes.io/} cluster \citep{2024SPIE13101E..3BA}. 

The OOQSManager component manages the life cycle of the OOQS and is implemented as an ACS component. It is interfaced with the Telescope Control System (TCS) to receive start and stop commands. The OOQSMonitor and OOQSCommand ACS components monitor (green lines) and manage (blue lines) the data quality pipelines. The two parts share messages using the ZeroMQ\footnote{https://zeromq.org/} framework.

We implemented the data quality pipeline using the rta-dataprocessor (RTADP) framework \citep{bulgarelli_rtadp}. This framework aims to improve the implementation of real-time analysis pipelines in the context of gamma-ray observations, both with space and ground-based instruments. The RTADP framework flexibility allows us to satisfy the software requirements in several scenarios by developing high-throughput and low-latency pipelines. The OOQS analyses all the packets received from the ADAS to execute a list of quality checks that identify abnormal conditions that must be corrected and reported to the Operator. The data packets generated by the Cherenkov camera are: scientific data packets, calibration packets, variance packets, and housekeeping. Each Cherenkov camera comprises 37 Photon Detection Modules (PDMs) with 64 pixels. The main data quality analysis executed for each telescope are: 
\begin{enumerate}
\item Calculate the trigger number for the camera and each camera's PDM;
\item Check that the trigger rates are inside a predefined range;
\item Check that the pixel ADC values of each Cherenkov camera are inside a predefined range;
\item Sample the data to obtain one camera plot per second for the different data packets;
\item Calculate the ratio between the high-gain and low-gain of each camera PDM in the variance packets;
\item Check if the pointing deviation and the point spread function (PSF) size are inside the nominal range.
\end{enumerate}

The OOQS sends the results of these analyses to the Quality Archive and the Operator Human Machine Interface (HMI) through Kafka. The Operator visualizes quick-look results from the HMI during the observation. There will be one OOQS pipeline for each of the nine ASTRI Mini-Array telescopes to maximize availability and reduce single-point failures. 

\section{Conclusions}

The OOQS will execute real-time data quality checks during ASTRI Mini-Array observations to detect abnormal conditions that can negatively impact the observations. We designed a new software architecture for the OOQS to deploy it in the on-site data center. The OOQS ACS components that manage the OOQS lifecycle are deployed in a virtualization system. At the same time, the OOQS data quality pipeline is executed in a Docker container deployed in the Kubernetes cluster.  

The abnormal conditions detected by OOQS are used to correct in real-time the observations and optimize the telescope's duty cycle. The OOQS is interfaced with other ASTRI Mini-Array sub-systems to inform them if an abnormal condition is detected. 

The OOQS data quality results are stored in the Quality Archive, and some can be visualized by the Operator using the Operator Human Machine Interface.

\acknowledgements This work was conducted in the context of the ASTRI Project. We gratefully acknowledge support from the people, agencies, and organisations listed here: \href{http://www.astri.inaf.it/en/library/}{http://www.astri.inaf.it/en/library/}. This paper went through the internal ASTRI review process. 

\bibliography{P508}  


\end{document}